\documentstyle[referee]{mn}
\newcommand{\reference}{\bibitem}
\newcommand{\beq}{\begin{equation}}
\newcommand{\eeq}{\end{equation}}
\newcommand{\bey}{\begin{eqnarray}}
\newcommand{\eey}{\end{eqnarray}}

\newcommand{\mpc}{\,{\rm {Mpc}}}
\newcommand{\kpc}{\,{\rm {kpc}}}

\newcommand{\kms}{\,{\rm {km\, s^{-1}}}}

\newcommand{\vc}{V_c}

\newcommand{\Rd}{R_d}
\newcommand{\rd}{R_d}

\newcommand{\MB}{M_B}

\newcommand{\omnow}{\Omega_0}

\newcommand{\brv}{\beta_{\rm rv}}
\newcommand{\Qrv}{Q_{\rm rv}}
\newcommand{\brm}{\beta_{\rm rm}}
\newcommand{\Qrm}{Q_{\rm rm}}
\newcommand{\btf}{{\beta_{\rm tf}}}
\newcommand{\Qtf}{Q_{\rm tf}}
\newcommand{\betan}{\beta_{\rm n}}

\def\m2l{\Upsilon_{\rm tot}}
\def\bml{\beta_{\Upsilon}}
\def\ah{\alpha_h}
\def\ad{\alpha_d}
\def\av{\alpha_v}

\input epsf

\title[] 
{The Evolution of Galactic Disks}
\author[]
{Shude Mao, H.J. Mo and Simon D.M. White
\thanks{E-mail: (smao, hom, swhite)@mpa-garching.mpg.de} \\
	Max-Planck-Institut f\"ur Astrophysik
	Karl-Schwarzschild-Strasse 1, 85748 Garching, Germany}
\date{Accepted ........
      Received .......;
      in original form .......}

\pagerange{\pageref{firstpage}--\pageref{lastpage}}
\pubyear{1997}

\begin{document}
\maketitle
\label{firstpage}
\begin{abstract}
 We use recent observations of high-redshift galaxies 
to study the evolution of galactic disks
over the redshift range $0 < z\la 1$. The data are inconsistent 
with models in which disks were
already assembled at $z=1$ and have evolved only in luminosity
since that time.
Assuming that disk properties change with redshift as powers of
$1+z$ and analysing the observations assuming an Einstein-de Sitter
universe, we find that for given rotation speed, 
disk scalelength decreases with $z$ as $\sim (1+z)^{-1}$,
total $B$-band mass-to-light ratio decreases with $z$ 
as $\sim (1+z)^{-1}$, and disk luminosity (again in $B$) depends only
weakly on $z$. These scalings are consistent with current data
on the evolution of disk galaxy abundance as a function of size and
luminosity. Both the scalings and the abundance evolution are close
to the predictions of hierarchical models for galaxy formation.
If different cosmogonies are compared, the observed evolution in
disk-size and disk abundance favours a flat low-$\Omega_0$ universe
over an Einstein-de Sitter universe. 
\end{abstract}


\section {Introduction}

 The Hubble Space Telescope (HST) and large ground-based telescopes
can now provide photometric and kinematic data for normal galaxies 
out to redshifts of about one.  Such observations are providing new results on
the evolution of disk galaxies, showing that the population
of these objects changes substantially with cosmic time. For example, 
disks of a given size appear brighter in the past, or equivalently
disks of given luminosity appear smaller  
(e.g. Schade et al 1996; Lilly et al 1997 and references
therein). A traditional approach to interpreting such data
has invoked luminosity evolution, assuming that normal giant galaxies 
were already assembled by redshift unity and have evolved primarily 
in luminosity since that time. The purpose of this paper is to examine
this hypothesis in the light of recent data on disk sizes, 
luminosities and kinematics. We find 
it to be untenable if the data are taken at
face value; strong evolution in the structure
of disks appears to be required. The form
of this evolution is quite similar to that expected in recent models
for disk formation in hierarchical cosmologies.

In \S 2, we use recent local and high-redshift galaxy samples to 
fit for the values of the exponents in power-law representations
of the evolution of disk sizes, luminosities, rotation speeds and
abundances. In \S 3, we derive the scaling relations predicted for
disk properties by hierarchical models and we compare them with the 
observational relations. \S 4 presents some discussion of these results.

To avoid confusion, we transform all cosmology-dependent 
observational quantities to an assumed Einstein-de Sitter (EdS) 
universe with $h=0.5$ (where $h$ is the present Hubble constant
in units of $100\kms \mpc^{-1}$). We transform theoretical predictions
in this way also, even when they refer to a different cosmology.

\section {Observational Facts}

To study the photometric and kinematic evolution of galaxies,
we obviously need both high-redshift and local samples. For the 
high-redshift data, we use primarily
the Canada-France Redshift Survey (hereafter CFRS, 
see Lilly et al 1995; Schade et al 1996) and 
the KECK kinematical study of Vogt et al (1997a,b). Ideally, we would like
to have a local galaxy sample that has both photometric and kinematic 
information. However, at present this is not yet available. We therefore
make use of several samples (Kent 1985; Pierce \& Tully 1988, 1992; Courteau
1996, 1997; de Jong \& van der Kruit 1994, de Jong 1996).
We use simple power-law
parametrisations of the evolution
of observational relations (size-rotation speed, size-luminosity,
abundance-luminosity, Tully-Fisher, etc.) and we determine the evolution 
exponents {\it empirically} from the data.

\subsection{The $\Rd$-$\vc$ Relation}

We first study the relation between disk size and rotation
speed. The local sample is from Courteau (1996, 1997)
whereas the high-redshift data are taken from Vogt et
al (1997a,b). There are respectively 306 and 16 galaxies in these two samples.
Figure 1a shows the histogram of $\log \Qrv \equiv \log \Rd/\vc$,
where $R_d$ (in $\kpc$) is the disk scale-length 
and $\vc$ (in $\kms$) is the rotation speed. The
$\rd$ distribution at fixed $\vc$ is
expected to scale in proportion to $\vc$ with a coefficient
depending on the redshift (see \S 3); the marginal distribution of
$\rd/\vc$ should thus isolate the redshift dependence clearly.
The dashed line is for the high-redshift data while the solid line 
is for the local sample. Clearly $\Rd/\vc$ is 
smaller at high-redshift, i.e., for given $\vc$, high-redshift disks
are smaller (Vogt et al 1997a,b). According to a Kolmogorov-Smirnov (KS) 
test, the probability that these two samples are 
drawn from the same parent distribution is 
$1.5 \times 10^{-4}$. Clearly the assumption that disk sizes have
not changed since $z=1$ is inconsistent with the data. This excludes
models with pure luminosity evolution.
To quantify the size evolution, we adopt simple parametrisation,
$
\Qrv(z) = \Qrv(0) (1+z)^{\brv},
$
and evolve each high-redshift galaxy to the present day. 
We then compare the evolved
distribution of $\Rd/\vc$ to that of the local sample using the KS test.
The resulting probability is shown as the solid line in Figure 1d. One
can see that the most probable value of $\brv$
(corresponding to the largest KS probability) is
$\brv=-0.95$, with values outside  $-1.4< \brv<-0.54$ excluded at
95\% confidence.

\begin{figure}
\epsfysize=16cm
\centerline{\epsfbox{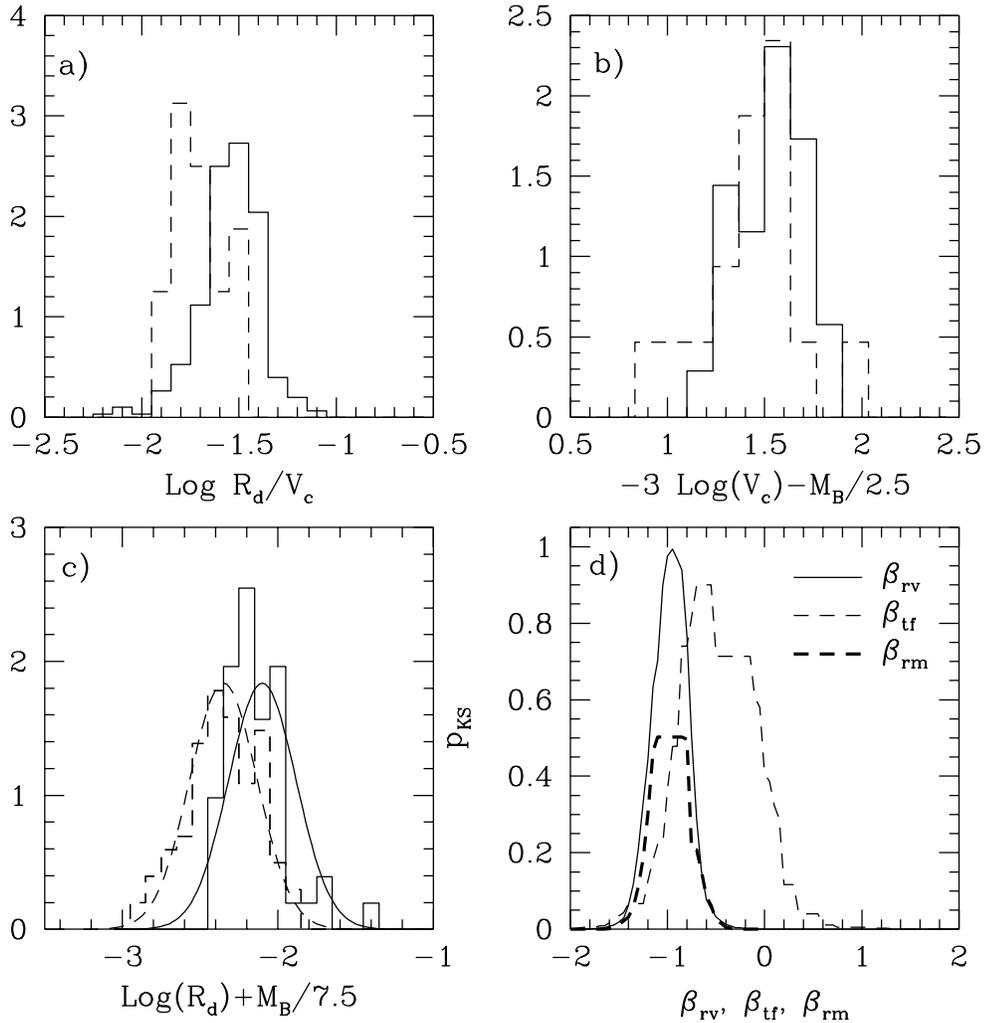}}
\caption{
Histograms for various physical quantities for local (solid lines) 
and high-redshift (dashed lines) samples. The area under all the histograms
is one.
(a) $\log \Qrv \equiv \log \rd/\vc$.
The local sample is from Courteau (1996, 1997) and has 306 galaxies.
The high-redshift sample from Vogt et al (1997a,b) has 
16 galaxies.
(b) $\log \Qtf \equiv -3\log (\vc)-M_B/2.5$.
The local sample is from Pierce \& Tully (1988) and has 26 galaxies.
The high-redshift sample is the same as in Fig. 1a.
(c) $\log \Qrm \equiv \log (\rd)+M_B/7.5$.
The local $r$-band sample is from Kent (1985) and has 51 galaxies. We
have converted it to B using $B-r=0.9$ (Kent 1984).
The high-redshift sample from the CFRS survey (Schade et al 1996) has
103 galaxies. The solid and dashed curves are normal fits
to the local and high-redshift data.
(d) KS test probability as a function of evolution exponents for the three
physical quantities plotted in panels a, b and c (see text).
}
\end{figure}

\subsection{The Tully-Fisher Relation}

Vogt et al (1997a,b) have studied the evolution of the Tully-Fisher relation
with redshift. They obtained kinematic data for high-$z$
disks using KECK spectroscopy and compared them to the local 
$B$-band Tully-Fisher relation of Pierce \& Tully (1988, 1992).
We have converted their quoted data to correspond to an assumed 
EdS cosmology with $h=0.5$ (see \S 3). 
Unlike the $R_d$-$V_c$ relation discussed above,
the comparison between the high- and low- redshift 
Tully-Fisher relations suffers from the uncertainties in the 
relative calibration of the various  photometric systems.
We follow Vogt et al in comparing 
to the Pierce \& Tully sample even though it contains only 26
galaxies in the Ursa Major cluster, but we note that the
Tully-Fisher zero-point of this cluster has a significant systematic
uncertainty (c.f. Giovanelli et al 1997).
We define the quantity $\Qtf \equiv L/\vc^3$,
where $L$ denotes galaxy luminosity. Since $L$ is roughly
proportional to $\vc^3$,
the distribution of $\Qtf$ can be used to measure evolution of the
zero-point of the Tully-Fisher relation.
Histograms of $\log \Qtf = -M_B/2.5-3 \log \vc$ are
shown in Fig. 1b. Here $M_B$ is the B-band absolute magnitude of the
galaxies. It is clear that the data are
consistent with no zero-point evolution, in agreement with the
conclusion of Vogt et al (1997a,b) and Hudson et al (1997).
A KS test shows the no-evolution assumption to be acceptable.
To put limits on evolution of the Tully-Fisher
relation, we set $\Qtf(z)=\Qtf(0)
(1+z)^{\btf}$ and perform KS tests on the ``evolved'' distributions
as before. The constraints on $\btf$ are indicated by the thin dashed line in
Figure 1d. With the current data, the evolution index is
not well constrained, only values outside $-1.44<\btf<0.36$ can be
excluded with better than 95\% confidence.   
More data, particularly at high-redshift, are needed to tie down 
the evolution of Tully-Fisher relation, and it is important to ensure
consistent photometric systems at high- and low-redshift. If we combine
the $\rd$-$\vc$ and Tully-Fisher relations, we can
estimate the evolution of the total $B$-band mass-to-light
ratio of the visible regions of galaxies:
$M/L \propto \vc^2 \rd/L = \Qrv/\Qtf \propto (1+z)^{\brv-\btf}$. For
$\btf=0$, this gives $M/L \propto 
(1+z)^{-1.0}$. However, if the most likely
value of $\btf$ is used, then $M/L \propto (1+z)^{-0.5}$.
The exponent here is quite uncertain as a
result of the uncertain evolution of the Tully-Fisher relation.

\subsection{The $\Rd$-$\MB$ Relation}

Next we examine the evolution of $R_d$-$M_B$ relation.
We compare the local (volume-limited) sample of Kent (1985) with the 
high-redshift data of Schade et al (1996).
We have checked that the independent local (diameter-limited)
sample of de Jong (1996) roughly agrees with that of Kent, when 
the different selection functions are properly taken into account.
Figure 1c shows histograms of the quantity 
$\log \Qrm \equiv \log(R_d)+M_B/7.5$ for the local and
high-redshift samples. Notice that for a tight Tully-Fisher relation
of the form $L \propto \vc^3$, $\Qrm$ should
be distributed similarly at each redshift to the $\Qrv$ discussed in \S 2.1. 
For a no-evolution model, the high- and low-redshift histograms
would coincide. Instead, one clearly sees that disks of given luminosity
are smaller at earlier times, or equivalently, galaxies of given size
are brighter. A KS test reveals that the probability that these two samples 
are drawn from the same distribution is  less than $4.9\times 10^{-6}$. 
Parametrising evolution of the size-luminosity relation as
$\Qrm(z) = \Qrm(0) (1+z)^{\brm}$, we estimate $\brm$ as before using a
KS test. The result is shown by the thick dashed
line in Figure 1d. The most probable value is
$\brm=-1.0$ and values outside $-1.35<\brm<-0.53$ can be excluded with
95\% significance. This result agrees with Schade et al (1996) who
found that the luminosity of disks of given size increases
approximately as $(1+z)^3$ in the CFRS. If we use only the big disks 
as defined in Lilly et al (1997), then we obtain $\brm \sim -0.5$.
This indicates that there might be some differential evolution
in the $\rd-M_B$ relation for small and big disks. Unfortunately,
current data do not allow a more detailed analysis of this evolution.

\subsection{Disk Size Functions}

Schade et al (1996) studied the redshift dependence of the abundance of 
disk galaxies as a function of size. By analogy with the luminosity
function this quantity is referred to as the ``size function''.
They found that the abundance of big, bright disks 
(scalelengths larger than $5\kpc$ and absolute magnitudes brighter than 
$-20$) is similar at $z\sim 0.3$ and at $0.75$, whereas
the abundance of small, bright
disks (scalelengths less than $4\kpc$ but $M_B<-20$) is $5 -10$ times
higher at the higher redshift. A comparison of
more recent HST imaging of CFRS and LDSS galaxies with the local
sample of De Jong et al (1996) confirms and strengthens this result
(Lilly et al 1997). In addition, if all galaxies to the CFRS limit are
included, the size functions for low- and high-redshift samples are similar.
These authors interpret their results as suggesting that the
abundances and sizes of big disks have not changed significantly since
$z=1$; the observed evolution reflects changes in disk luminosity 
caused by changes in star formation rate. This interpretation conflicts with
our earlier conclusions, and in this subsection we test whether the 
observed size functions are also consistent with evolution of the kind
inferred above.

\begin{figure}
\epsfysize=16cm
\centerline{\epsfbox{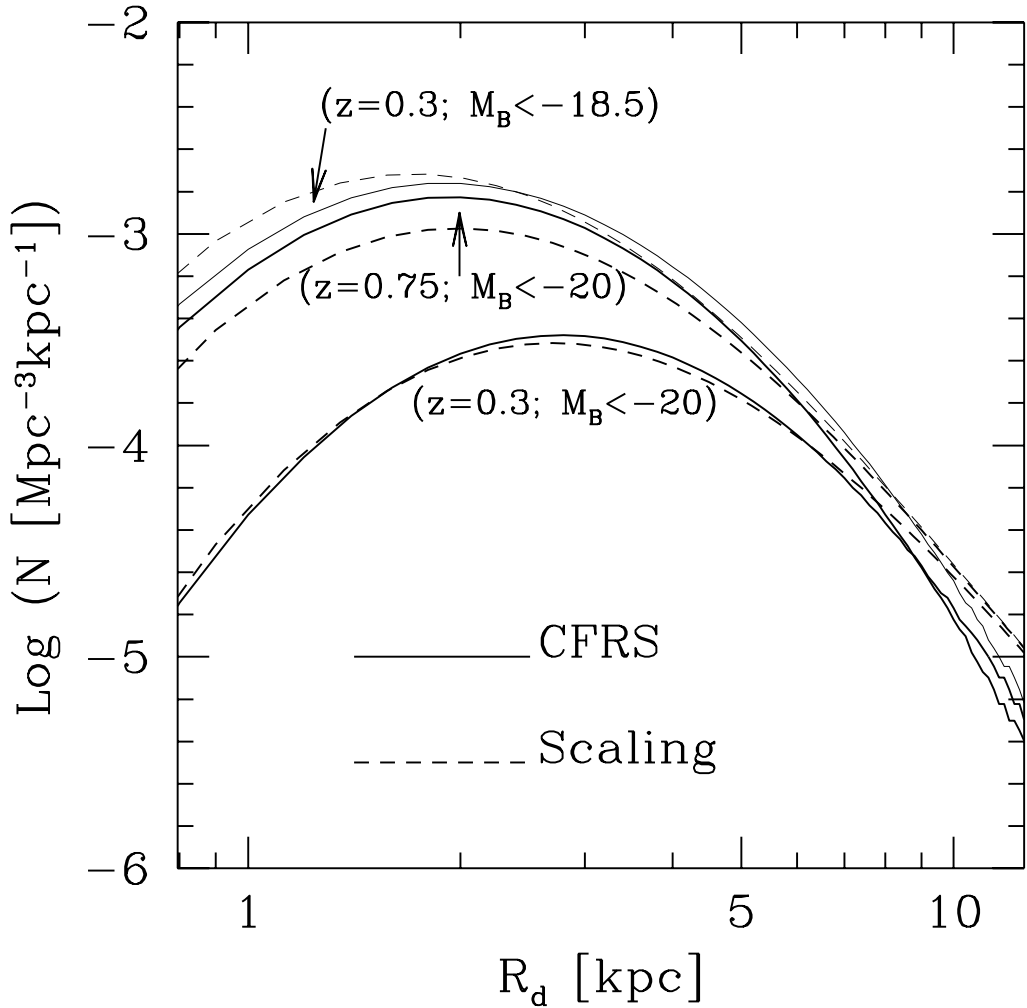}}
\caption{The size functions  
for galactic disks.
The solid curves show results derived directly from the
CFRS, whereas the dashed curves show the model of
equation (\ref{nrd}) for $\brv=-1.0$, $\betan=3.0$,
$\gamma=-4$ and $N_\star=1.0\times 10^{-3}{\rm Mpc}^{-3}$.
The lower pair of curves is for galaxies with $M_B<-20$
at $z=0.3$; the middle pair for galaxies with $M_B<-20$
at $z=0.75$; the upper pair for galaxies with $M_B<-18.5$
at $z=0.3$.}
\end{figure}

We start with the observed luminosity functions
for disk galaxies at various redshifts. Using an empirical Gaussian
fit to the distributions of $\log(R_d)+M_B/7.5$
in Fig. 1c, and assuming that for given
$M_B$ disk size varies as
$R_d\propto (1+z)^{-1}$ (see \S 2.3), 
we can transform the luminosity function into a size 
function at each redshift. The solid curves in Figure 2 show
the size functions derived from the luminosity functions
given by Lilly et al (1995) for blue galaxies (which are 
mostly late-type galaxies) in the CFRS. Results are shown for two
absolute magnitude limits $M_B=-20$ and $M_B=-18.5$:
the former is approximately the magnitude limit for the high-redshift
sample (with median redshift 0.75)
 whereas the latter is that for the low-redshift sample (with median
redshift 0.3). We see that the abundance of disks with $\rd>4\kpc$
hardly changes with redshift while the abundance of
smaller galaxies changes substantially. Not surprisingly, this agrees
well with the conclusions of Schade et al (1996) and Lilly et al (1997)
since both our evolution rate and our distribution of $R_d$ for given 
$M_B$ were derived from the CFRS data. 

We now compare these size functions with a model based on
the kind of evolution inferred in previous sections.
Let us approximate the (comoving) abundance of galaxies 
with rotation speed $\vc$ at redshift $z$ as
\beq\label{nvcz}
N(\vc, z)d\vc=N_\star (1+z)^{\betan} 
\left({\vc\over V_\star}\right)^{\gamma}{d\vc\over V_\star},
\eeq
where $V_\star$ is some fiducial rotation speed, 
$N_\star$ is the abundance of $V_\star$ 
galaxies, $\betan$ parameterises the abundance 
evolution, and $\gamma$ parameterises the shape of the 
rotation speed function. Note that this is a {\it minimal} 
parametrisation, which can be valid only over limited ranges 
of $\vc$ and $z$. In practice we are interested in values of $\vc$ 
similar to those of $L_\star$ galaxies. If, as observations
suggest, disks show a reasonably tight Tully-Fisher relation at each
redshift, we can use the parametrisation of \S 2.2 to convert
this to a luminosity function,
\beq\label{nLz}
N(L,z)dL = {1\over 3} N_\star (1+z)^{\betan} 
\left({L\over L_\star(z)}\right)^{(\gamma-2)/3}{dL\over L_\star(z)}
\propto (1+z)^{\betan-(\gamma+1)\btf/3}L^{(\gamma-2)/3}dL,
\eeq
where $L_\star(z)=(1+z)^{\btf} L_{\star ,0}$ and $L_{\star ,0}$ is the
luminosity corresponding to $V_\star$ at $z=0$.

In \S2.1 we found that the mean size-rotation speed relation can be 
well represented by $R_d\propto \vc(1+z)^{\brv}$. The scatter
in $\ln R_d$ about this relation is approximately normal with an {\it
rms} of $\sigma\approx 0.5$ (cf. Fig. 1c).
This allows us to calculate the size function
for galaxies with absolute $B$-magnitude 
brighter than $M_{\rm lim}$ and so rotation speed greater than
$(L_{\rm lim}/L_\star)^{1/3}V_\star$. We find
\bey\label{nrd}
N(R_d, z)dR_d&=&N_\star (1+z)^{\betan} 
\left[{R_d\over R_\star(z)}\right]^{\gamma}
{dR_d\over R_\star(z)}
\nonumber\\
&&\times \int_{-\infty}^{x_{\rm max}}
{1\over \sqrt{2\pi}\sigma}
\exp\left[-{x^2\over 2\sigma ^2}
-(1+\gamma)x\right] dx;
\eey
\beq
x_{\rm max}=[\ln(10)/7.5]\left(M_{\rm lim}-M_\star\right)
+\ln [R_d/R_\star(z)],
\eeq
where $M_\star$ is the absolute magnitude corresponding to 
$L_\star$, and
$
R_\star(z)=R_0 (1+z)^{\brv} 
$
is the median value of disk scalelength for galaxies   
with $\vc=V_\star$ at redshift $z$. From
the solid histogram shown in Fig. 1a we have 
$(R_0/{\rm kpc})\approx 0.03 (V_\star/\kms)$.
For $M_{\rm lim}\ga M_\star$ and 
$R_d\gg R_\star (z)$, the integration in 
equation (\ref{nrd}) depends only weakly on $R_d$, and
the size function scales with $R_d$ and $z$ as
\beq
N(R_d, z)dR_d\propto (1+z)^{\betan-(1+\gamma)\brv}
R_d^{\gamma}dR_d.
\eeq
Thus, the observational result that the comoving number density 
of large and bright  disks does not change significantly with redshift
requires $\betan \sim \brv(1+\gamma)$. 

The interpretation of Schade
et al (1996) and Lilly et al (1997) corresponds to $\betan=\brv=0$ and
$\btf=3$ -- no evolution in abundance or size but strong evolution
in luminosity. In contrast, our earlier analysis suggested
$\brv\sim -1$, implying $\betan \sim -(1+\gamma)$. At
luminosities around $L_\star$ the slope of observed luminosity
functions is near $-2$ which, according to equation (\ref{nLz}),
corresponds to $\gamma\sim -4$. Consistency then requires $\betan\sim 3$.
In Fig.~2 we show the predictions of equation (\ref{nrd}) for 
$\brv=-1.0$, $\betan=3.0$, $\gamma=-4$, $N_\star =1.0\times 10^{-3}
{\rm Mpc}^{-3}$ and $M_{\rm lim}=-20$.
These agree well both with the size functions derived from
the luminosity functions of Lilly et al (1995) and with the direct 
observational estimates of Schade et al (1996). The figure also shows
predictions for disks at $z=0.3$ with $M_B<-18.5$.
The size function of this low-redshift sample
is similar to that of the high-redshift sample limited at
$M_B=-20$ as found by Lilly et al (1997) in their observational samples.
We conclude that the observed 
size functions are indeed compatible with a model
in which disks of given rotation speed have a size which varies as
$(1+z)^{-1}$ and an abundance which varies as
$(1+z)^3$. If the redshift dependence of the Tully-Fisher relation 
is weak, then at fixed luminosity the comoving density in this
model evolves approximately as $(1+z)^3$ (see equation [\ref{nLz}]).
This agrees with the
behaviour which Lilly et al (1996) inferred from an analysis of the 
comoving $B$-band luminosity density.

\section {Model Scalings}

The analysis of the last section shows that current data are 
inconsistent with the hypothesis that disk galaxies were already
assembled at redshift unity and have only evolved in luminosity
since that time. The current section compares with the 
predictions of hierarchical models for galaxy formation, based on the
work reported in Mo, Mao \& White (1997). 
To clarify the relevant scalings, it is sufficient to treat galaxies
as exponential disks embedded in isothermal halos; for simplicity we
neglect both the disk mass and the halo core radius.
The luminosity and size of the disk then scale as
\beq \label{l-rd}
L \propto {1 \over \m2l(z)} \vc^3 {H_0 \over H(z)}, ~~
\rd \propto \lambda {\vc \over H_0}{H_0 \over H(z)},
\eeq
where $\m2l$ is the total mass-to-light ratio within the {\it halo
virial radius} and $\lambda$ is the dimensionless spin parameter.
For given circular velocity, disk size depends on redshift only
through the Hubble constant $H(z)$, 
while the luminosity has an additional dependence through 
the mass-to-light ratio. If we make the approximations 
$\m2l(z) = \m2l(0) (1+z)^{\bml}$ and $H(z) = H_0 (1+z)^{\beta_{\rm h}}$, 
then from the definitions, $\Qtf \equiv L/\vc^3$
and $\Qrv \equiv \rd/\vc$, we have 
\beq
\Qtf(z) = \Qtf(0) (1+z)^{-\bml-\beta_{\rm h}}, ~~~~ 
\Qrv(z)=\Qrv(0)
(1+z)^{-\beta_{\rm h}}.
\eeq
Thus we see that $\beta_{\rm h}$ is equal to $\brv$ as defined
above. In an EdS universe, $\beta_{\rm h}=3/2$, so this
simple model predicts that for given
circular velocity, the mean size of disks should vary as $(1+z)^{-3/2}$.
This is somewhat too steep to be compatible with the data analysed
in \S 2.1 (cf. Figure 1d). 

In order to compare the data with 
predictions for other cosmologies, we can either 
reanalyse the observations assuming the cosmology 
under consideration, or transform the model predictions
into those expected when the data are analysed {\it assuming}
an EdS universe. Here we adopt the second approach. Cosmology enters
our problem through the Hubble constant, the angular size and 
luminosity distances and the comoving volume.
For moderate redshifts ($0<z<1$), the ratios of these
quantities in a general cosmology to those in an EdS universe can be
approximated by power-laws in $1+z$:
\beq \label{ratio}
{H(z) \over \widetilde{H}(z)} = (1+z)^{\ah}, ~
{d_A(z) \over \widetilde{d_A}(z)} = (1+z)^{\ad}, ~
{dv/dz \over \widetilde{dv/dz}} = (1+z)^{\av},
\eeq
where the tilded quantities refer to the EdS cosmology;
$d_A$ is the angular diameter distance and $dv/dz$ is the
differential comoving volume. The indices $\ah, \ad$ and $\av$ can be
approximated by
\beq \label{flat}
\alpha_h = -1.5 (1-\omnow^{0.90}),~
\alpha_d = 0.78 (1 - \omnow^{0.67}),~
\alpha_v = 3.1 (1 - \omnow^{0.65}), 
\eeq
for flat cosmologies with $\Omega_0+\Lambda=1$,
and by
\beq \label{open}
\alpha_h = -0.5 (1-\omnow^{0.65}), ~
\alpha_d = 0.35 (1-\omnow^{0.90}), ~
\alpha_v = 1.2 (1-\omnow^{0.80}), 
\eeq
for open cosmologies with $\Omega_0\le 1$ and $\Lambda=0$.
Using the definitions of $\Qrv$, $\Qrm$ and $\Qtf$ together with
equations (\ref{l-rd}) and
(\ref{ratio}) we have
\beq
\brv = - \ad - \ah - {3 \over 2}, ~
\btf = -2\ad - \ah - \bml - {3 \over 2}, ~
\brm = \brv - {\btf \over 3}.
\eeq
These allow the predicted evolution exponents for any cosmological 
model to be compared directly with the observed exponents
estimated assuming an EdS universe.  
For example, a flat cosmology with $\Omega_0=0.3$ and $\Lambda=0.7$ has
$\ah=-1.0, \ad=0.43$ and $ \av=1.68$. Hence $\brv = -0.93$ 
and $\btf=-1.36-\bml$. If there is no evolution in the Tully-Fisher 
zero-point (cf. \S2.2), then $\bml=-1.36$, and $\brm=-0.93$, in reasonable
agreement with the exponents inferred from the data in \S 2.

The values of $\betan$ and $\gamma$ characterizing the 
number-density evolution of disks (cf. equation [\ref{nvcz}]) 
can be estimated from the
Press-Schechter formalism (Press \& Schechter 1974). This provides
simple formulae for the comoving abundance of haloes as a
function of redshift and of halo mass $M_{\rm h}$ or circular
velocity $\vc$ (e.g. White \& Frenk 1991). For currently popular 
cosmogonies, the slope of the abundance
curves for $\vc$ values corresponding to the rotation speeds of observed
disks implies $\gamma\sim -3.5 \to -3.8$, while the evolution of
the abundance at fixed $\vc$ in this range gives $\betan\sim 
1.3\to 1.7$ for CDM-like models with $\Omega_0=1$ and
$\betan\sim 2.9 \to 3.3$ for a flat model with $\Omega_0=0.3$
and $\Lambda=0.7$. The larger $\betan$ in the latter
model is due to the volume correction 
to the EdS universe (cf. equation \ref{flat}).
These values are quite similar to those inferred empirically
in \S 2, particularly for the flat model with low $\Omega_0$.
Thus the observed evolution both in abundance and in
systematic parameter relations seems compatible with the expectations
of hierarchical formation theories.

\section{Discussion}

We have shown that current data appear inconsistent with models 
in which disk galaxies are already assembled by redshift one and
have evolved only in luminosity since that time.
Assuming that disk properties vary as powers of
$1+z$ and analysing them under the assumption of an EdS
universe, we find that for given rotation speed, 
disk scalelength varies as $\sim (1+z)^{-1}$, 
total mass-to-light ratio (in the $B$-band) 
as $\sim (1+z)^{-1}$, disk luminosity (in $B$-band) is almost 
independent of $z$, and disk abundance varies
as $\sim (1+z)^3$. These scalings agree well with the predictions 
of hierarchical models for flat low-density universes but are
marginally weaker than expected in an Einstein-de Sitter universe. 
Our results for
the evolution of the size, luminosity and abundance of disk galaxies
can be compared to those of Bouwens, Broadhurst \& Silk
(1997) who modeled the counts of faint galaxies as a
function of apparent magnitude, colour and size in the Hubble Deep 
Field (Williams et al 1996). Their preferred model corresponds to 
$\betan \sim 3.5$ and 
$\brm \sim -1.5$, similar to the values derived here.

In the tests described in this paper, our own preferred model 
is distinguished from the simpler
hypothesis proposed by Schade et al (1996) and Lilly
et al (1997) only because we have included kinematic data.
Our results therefore depend critically
on the high-redshift observations of Vogt et al (1997a,b).
The apparent evolution of the size-rotation speed
relation is inconsistent with pure luminosity evolution,
which also requires stronger luminosity evolution than can be
reconciled with the weak redshift-dependence of the
Tully-Fisher relation. The Vogt et al sample is quite small and 
the comparison of its Tully-Fisher zero-point to that of local samples
is uncertain. Further kinematic and photometric data for high-redshift 
galaxies are necessary to confirm
the conclusions we have reached. Only with the inclusion of kinematic
information is it easy to separate evolution in the mass and
size of galaxies from evolution in their stellar mass-to-light ratio.

\section*{Acknowledgments}

We are grateful to Nicole Vogt and Stephane Courteau for providing us data
in electronic form.
This project is partly supported by
the ``Sonderforschungsbereich 375-95 f\"ur Astro-Teilchenphysik'' der
Deutschen Forschungsgemeinschaft.

{}

\newpage

\bsp
\label{lastpage}
\end{document}